\documentclass[twoside,twocolumn]{article}

\usepackage{blindtext} 

\usepackage[sc]{mathpazo} 
\usepackage[T1]{fontenc} 
\usepackage{microtype} 

\usepackage[english]{babel} 

\usepackage[hmarginratio=1:1,top=32mm,columnsep=20pt]{geometry} 
\usepackage[hang, small,labelfont=bf,up,textfont=it,up]{caption} 
\usepackage{booktabs} 

\usepackage{lettrine} 

\usepackage{enumitem} 
\setlist[itemize]{noitemsep} 

\usepackage{abstract} 

\usepackage{titlesec} 
\renewcommand\thesection{\Roman{section}} 
\renewcommand\thesubsection{\roman{subsection}} 
\titleformat{\section}[block]{\large\scshape\centering}{\thesection.}{1em}{} 
\titleformat{\subsection}[block]{\large}{\thesubsection.}{1em}{} 

\usepackage{fancyhdr} 
\usepackage{titling} 

\usepackage{hyperref} 

\usepackage[sorting=none]{biblatex}
\usepackage{relsize}
\usepackage{listings}
\usepackage{listings-rust}
\usepackage{xcolor}
\usepackage{float}
\usepackage{graphicx}
\usepackage{siunitx}
\usepackage{tabularx}
\usepackage{multirow}
\usepackage{balance}

\definecolor{codegreen}{rgb}{0,0.6,0}
\definecolor{codegray}{rgb}{0.5,0.5,0.5}
\definecolor{codepurple}{rgb}{0.58,0,0.82}
\definecolor{backcolour}{rgb}{0.95,0.95,0.92}

\lstdefinestyle{mystyle}{
    backgroundcolor=\color{backcolour},
    commentstyle=\color{codegreen},
    keywordstyle=\color{magenta},
    numberstyle=\tiny\color{codegray},
    stringstyle=\color{codepurple},
    basicstyle=\ttfamily\scriptsize,
    breakatwhitespace=false,
    breaklines=true,
    captionpos=b,
    keepspaces=true,
    numbers=left,
    numbersep=5pt,
    showspaces=false,
    showstringspaces=false,
    showtabs=false,
    tabsize=2
}

\newcommand{\lj}{soc-LiveJournal1}
\newcommand{\twitter}{twitter\_rv}
\newcommand{\uk}{uk-2007-05}

\pretitle{\begin{center}\Huge\bfseries} 
\posttitle{\end{center}} 
\title{COST of Graph Processing Using Actors}
\author{%
\textsc{Ronak Buch}\\[1ex] 
\normalsize University of Illinois at Urbana-Champaign \\ 
\normalsize \href{mailto:rabuch2@illinois.edu}{rabuch2@illinois.edu} 
}
\date{} 
\renewcommand{\maketitlehookd}{%
\begin{abstract}
  Graph processing is an increasingly important domain of computer science, with
  applications in data and network analysis, among others. Target graphs in
  these applications are often large, leading to the creation of ``big data''
  systems designed to provide the scalability needed to analyze these graphs
  using parallel processing. However, researchers have shown that while these
  systems often provide scalability, they also often introduce overheads that
  exceed the benefits they provide, sometimes resulting in lower absolute
  performance than even simple serial implementations.

  This report studies the viability and performance of the actor model to
  implement scalable concurrent programs to perform common graph computations.
  We show that relatively simple actor-based implementations outperform both
  dedicated graph processing systems and the benchmark serial implementations.
\end{abstract}

}

\begin{document}

\maketitle


\section{Introduction}
\label{sec:intro}

Scalability is often heralded as the most important property of a ``big data''
software system. In~\cite{McSherryCOST}, \citeauthor*{McSherryCOST} show that
this metric is frequently misleading when not coupled with performance
comparisons against an appropriate baseline. Providing decreased runtime as core
count increases is certainly a desirable property of a well-engineered parallel
system, but, in the end, absolute performance is paramount from the perspective
of a user, trumping scalability.

\citeauthor*{McSherryCOST} introduce a metric they term COST, or
\textbf{C}onfiguration that \textbf{O}utperforms a \textbf{S}ingle
\textbf{T}hread. This metric measures the amount of hardware resources required
before a particular system beats the runtime of a reasonably designed single
threaded solution. In particular, they found that several parallel data
processing systems from the literature were outperformed by a simple serial
program for various common graph processing tasks, even when the parallel
systems were given over 100x the computational resources. This analysis suggests
that these big data systems achieve their scalability by parallelizing overhead
that they \textit{themselves} have introduced rather than providing a useful
speedup relative to a fair baseline.

While some problem domains have inherent algorithmic or dataset related
impediments to achieving performance improvements with parallel execution, such
issues are not present here. Datasets are large, providing the scale to
decompose across several processors, and algorithms are generally data parallel
without fine-grained global synchronization or dependencies, only requiring
interactions within a local neighborhood. Given this potential, well-engineered
parallel implementations should be able to provide speedups over the serial
versions.

Taking inspiration from the findings of the COST work, in this report, we test
this supposition. We implement parallel versions of two common graph algorithms
using an actor-based framework and compare performance and programmabilty
against a well-established existing parallel graph processing system and the
baseline serial implementations.




\section{Approach}
\label{sec:approach}

In this section, we introduce and describe the ideology and principles of the
actor model and the specific framework used in our implementations, Charm++. We
also discuss the graph computations performed by our implementations.

\subsection{Actor Model}

The actor model~\cite{actor} provides a conceptual framework to reason about and
design concurrent computation. In this model, the computation is decomposed into
several entities called \textit{actors}. Each individual actor operates
independently from the others, with its own private state that no other object
can read or modify. Actors pass data and coordinate execution via sending
\textit{messages} to each other. Notably, this messaging is entity-centric (i.e.
an actor is specified as the destination of a message) as opposed to
processor-centric, as seen in the bulk synchronous parallel model or message
passing \`a la MPI. When an actor receives a message, it gains ownership over
the payload data. In response to a message, an actor can perform some
computation, send additional messages, spawn new actors, etc.

This encapsulation of state and restriction of interaction to only messages
reduces the complexity of implementing concurrent applications. Rather than having
to think about the global state of the application or multiple non-deterministic
threads of execution and their interaction with each other, the programmer
merely has to consider the received messages and local state on a given
individual actor when designing the logic of the program.

Using actors improves the safety of parallel applications as compared to
shared memory concurrency. Since state is private and cannot be concurrently
accessed or altered, the actor model prevents memory access order race
conditions and obviates the need for lock-based synchronization.

Programs expressed in the actor model can naturally be executed in parallel due
to the inherently concurrent semantics of actor computation. Further, the
messaging semantics allow execution to be fully distributed, since there is no
need for shared state.

\subsubsection{Concrete Implementations}

Due to these properties, concrete implementations of the actor model are popular
in the field of distributed systems, with examples such as ActorFoundry, Akka,
and Erlang, among others. These implementations often exploit the properties of
actor-based computation to add additional features such as fault tolerance and
migratability.

However, while these languages and libraries provide the many benefits of the
actor model to users, a common concern is that they are focused more on
expressivity and programmability than on pure raw performance. The need to do
location management, message delivery, garbage collection, etc. adds overhead
relative to more primitive, lower-level forms of writing parallel programs.
Given that we are interested primarily in a performance comparison, it is
important that our actor-based implementations utilize a platform designed with
performance in mind, such as Charm++.

\subsubsection{Charm++}

Charm++~\cite{charm-zenodo}~\cite{Kale2013} is an adaptive, asynchronous
task-based parallel programming framework based on the actor model and focused
on high performance computing applications. The core entities in a Charm++
program are objects called \textit{chares}, which are essentially actors. Chares
hold private state and generally communicate with each other using messages. The
Charm++ runtime system automatically measures the load of chares, and coupled
with the message delivery semantics of actors, can use these measurements to
migrate chares to automatically balance load between processors.

Both the runtime of Charm++ and applications that use it are written in C++,
meaning there is no garbage collection and the overheads are that of a low-level
systems programming language.

While one may use Charm++ to develop applications that completely adhere to the
semantics of the actor model, it allows users to break some of these constraints
in the interest of performing performance optimizations. For example, users can
create shared buffers between chares to more efficiently share data as compared
to messaging. These boundary violations must be done carefully and can often
lead to degraded performance if not done well, but they are often needed in the HPC
domain where performance is sacrosanct.

Additionally, Charm++ offers some programming niceties such as quiescence
detection, automatic message aggregation, and the ability to do bulk creation of
actors in collections called ``chare arrays'', which allow actors to also be
referenced via an index rather than just an address.

\subsection{Computations}

We implement two common graph computations, PageRank and connected component
detection via label propagation.

\subsubsection{PageRank}

PageRank~\cite{pagerank} is a method for ranking vertices in a directed graph by ``link
popularity.'' Famously, it was developed by the founders of Google and served as
the original algorithm underpinning the results of their eponymous web search
engine.

PageRank is an iterative algorithm that maintains a rank for each vertex in the
graph. During each iteration, the rank of a vertex is dampened by a given factor
$\alpha$, divided by its out-degree $d$, and sent to each of its outgoing
neighbors. The new rank of a vertex is computed by adding all of the
contributions it receives from its incoming neighbors to $1 - \alpha$.

The original implementation from the COST paper is shown in
Listing~\ref{lst:cost_pr} and the Charm++ implementation in
Listing~\ref{lst:charm_pr} (declarations and initialization have been condensed
or omitted for space).






\begin{lstlisting}[language=Rust,caption=Serial PageRank,float,label=lst:cost_pr]
fn PageRank20(graph: GraphIter, alpha: f32) {
  let mut a = vec![0f32; graph.nodes()];
  let mut b = vec![0f32; graph.nodes()];
  let mut d = vec![0f32; graph.nodes()];

  graph.map_edges(|x, y| { d[x] += 1; });
  for iter in 0..20 {
    for i in 0..graph.nodes() {
      b[i] = alpha * a[i] / d[i];
      a[i] = 1f32 - alpha;
    }
    graph.map_edges(|x, y| { a[y] += b[x]; })
  }
}
\end{lstlisting}

\begin{lstlisting}[language=C++,caption=Charm++ PageRank,float,label=lst:charm_pr]
// Driver function, only runs on 0th chare
void runpagerank(float alpha) {
  for (int i = 0; i < 20; i++) {
    // Call update on all chares in array
    thisProxy.update(alpha);
    CkWaitQD(); // Sleep until quiescence
    // Call iterate on all chares in array
    thisProxy.iterate();
    CkWaitQD();
  }
}

// Set up values for new iteration
void update(float alpha) {
  for (int i = 0; i < d.size(); i++) {
    b[i] = alpha * a[i] / d[i];
    a[i] = 1 - alpha;
  }
}

// Run PageRank iteration, send to neighbors
void iterate() {
  vector<vector<pair<int, float>>> outgoing;
  auto edgeIt = edges.begin();
  for (int i = 0; i < degs.size(); i++) {
    for (int j = 0; j < degs[i]; j++) {
      const auto dest = *edgeIt++
      outgoing[CHUNKINDEX(dest)].emplace_back(dest, b[i]);
    }
  }
  for (int i = 0; i < outgoing.size(); i++) {
    // Call addB on chare with index i
    thisProxy[i].addB(outgoing[i]);
  }
}

// Receive values from neighbor
void addB(vector<pair<int, float>> b_in) {
  for (const auto& entry : b_in) {
    const auto dest = entry.first;
    const auto value = entry.second;
    // base is first index on this chunk
    a[dest - base] += value;
  }
}
\end{lstlisting}

\subsubsection{Connected Components}

A connected component of a graph is a connected subgraph that is not part of any
larger connected subgraph, or, more formally, a subgraph $C$ of an undirected
graph $G$ such that every vertex in $V(C)$ is reachable from all vertices in
$V(C)$ and not reachable from any vertex in $V(G) - V(C)$.

There are many known algorithms to find connected components, but here we use
the \textit{label propagation} method~\cite{pregel} due to the suitability of
its neighborhood communication pattern for distributed computation. Label
propagation is an iterative algorithm that maintains a label for each vertex in
the graph, initially set to the unique index of the vertex. In each iteration,
the current label of each vertex is sent to each of its neighbors. Upon
receiving a candidate label from a neighbor, a vertex changes its label to the
candidate if the candidate is strictly less than its current label. This process
continues until an iteration occurs where no vertex changes its label. At
termination, the label of each vertex is equal to the smallest index of the
vertices in its component.

\section{Methodology}
\label{sec:methodology}

We compare our implementations to two references, the same serial
versions\footnote{https://github.com/frankmcsherry/COST} implemented in Rust as
used in the original COST paper, and the implementations provided by
GraphX~\cite{graphx}, the graph processing component of the Apache Spark data
analysis engine.

GraphX was one of the ``big data'' systems compared in the COST paper, and it either
approximately matched in performance or outperformed the other big data systems
used in that comparison. Here, we take it as a representative for the landscape
of data processing systems due to its popularity and past performance.
Additionally, on their website\footnote{https://spark.apache.org/graphx/},
GraphX claims to have ``Comparable performance to the fastest specialized graph
processing systems.''

We evaluate the performance of the implementations on three different real-world
input graphs:

\begin{itemize}
  \item \lj{}~\cite{snapnets} - Friendship network graph of the social media service LiveJournal
  \item \twitter{}~\cite{Kwak10www} - Following network graph of the social media service Twitter
  \item \uk{}~\cite{BoVWFI}~\cite{BRSLLP} - Hyperlink graph of web pages in the \texttt{.uk} domain
\end{itemize}

\begin{table}
  \centering
  \small
\begin{tabularx}{\linewidth}{l|rr}
\toprule
Name & Vertices & Edges \\
\midrule
  \lj{} & $4,847,571$ & $68,993,773$ \\
  \twitter{} & $61,578,415$ & $1,468,365,182$ \\
  \uk{} & $105,896,555$ & $3,738,733,648$ \\
\bottomrule
\end{tabularx}
\caption{Properties of Selected Graphs}
\label{tab:graphs}
\end{table}

The properties of these graphs are given in Table \ref{tab:graphs}. \twitter{}
and \uk{} are the same datasets used for evaluation in the original COST paper
and widely used in the benchmarking of big data systems, and \lj{} is a smaller
dataset added to ease testing during development of the new actor-based
versions.

Note that all of the chosen graphs are directed graphs. For use with label
propagation, input graphs are converted to undirected versions by adding a
reverse edge for each edge if it does not already exist in the edge set, meaning
that the graphs have more edges than shown in Table~\ref{tab:graphs} during
label propagation.

Experiments were conducted on the CPU partition of
Delta\footnote{https://delta.ncsa.illinois.edu/} at the National Center for
Supercomputing Applications. Each node of Delta in the CPU partition has two AMD
EPYC 7763 processors, with 128 cores across two sockets and 256 GB of memory.

Finally, note that all provided timings are only of the actual computation, the
time taken to ingest the graph from storage and perform other upfront
preparation and initialization are not included.

\section{Results \& Discussion}
\label{sec:results}

\subsection{Serial and GraphX Baselines}

\begin{figure}
  \includegraphics[width=\linewidth]{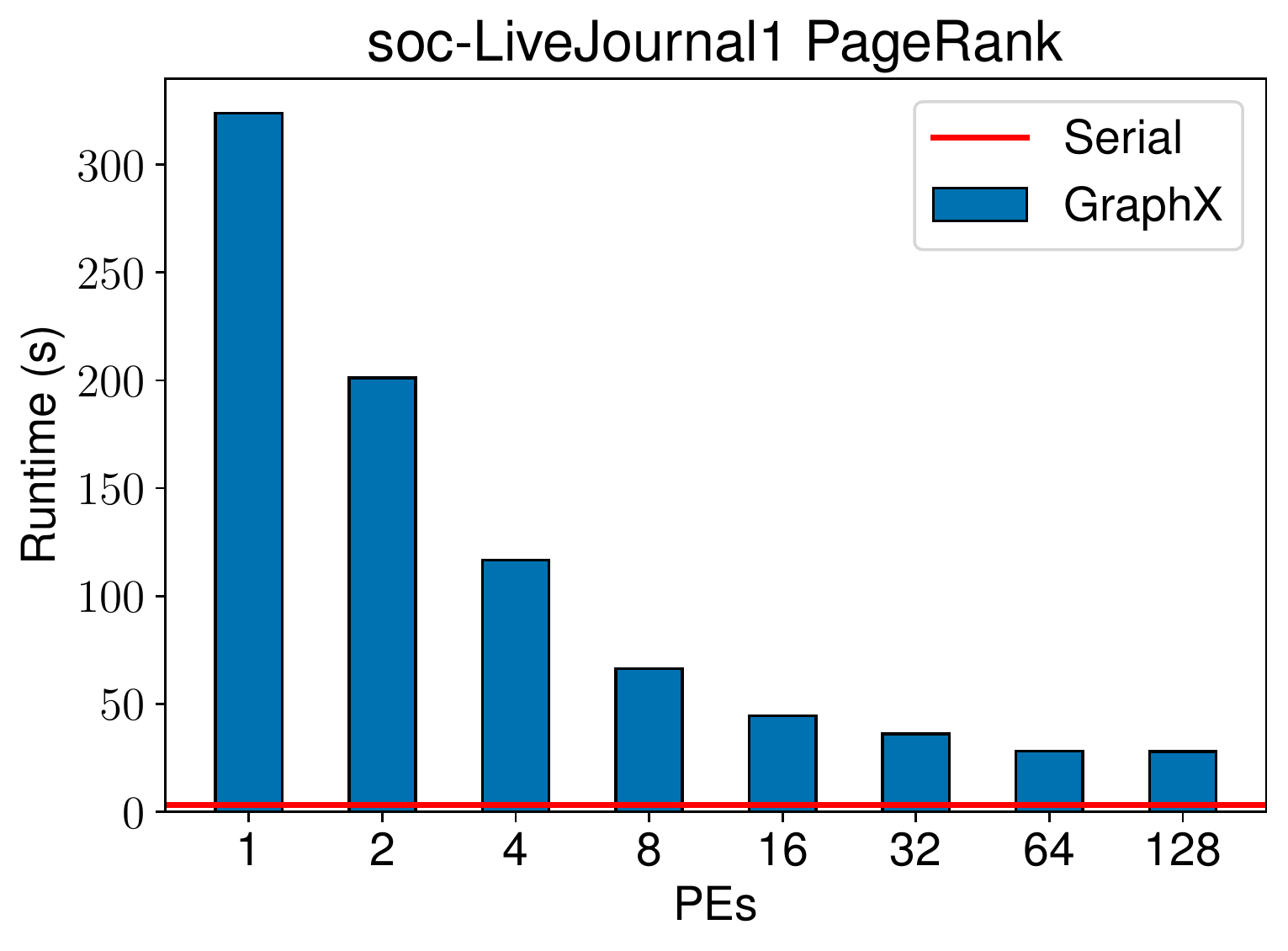}
  \caption{GraphX PageRank on \lj{}}
  \label{fig:graphx_pr}
\end{figure}

In order to establish a baseline for comparison, we first examine the GraphX and
serial implementations. Figure~\ref{fig:graphx_pr} shows the performance of
these baseline runs for PageRank with \lj{}. GraphX performance scales as the
number of processors (PEs) increases, but even with its fastest configuration of
128 PEs, it takes 27.9~\si{s} for 20 iterations of PageRank, while the serial
version takes merely 3.18~\si{s} on a single core.

\begin{figure}
  \includegraphics[width=\linewidth]{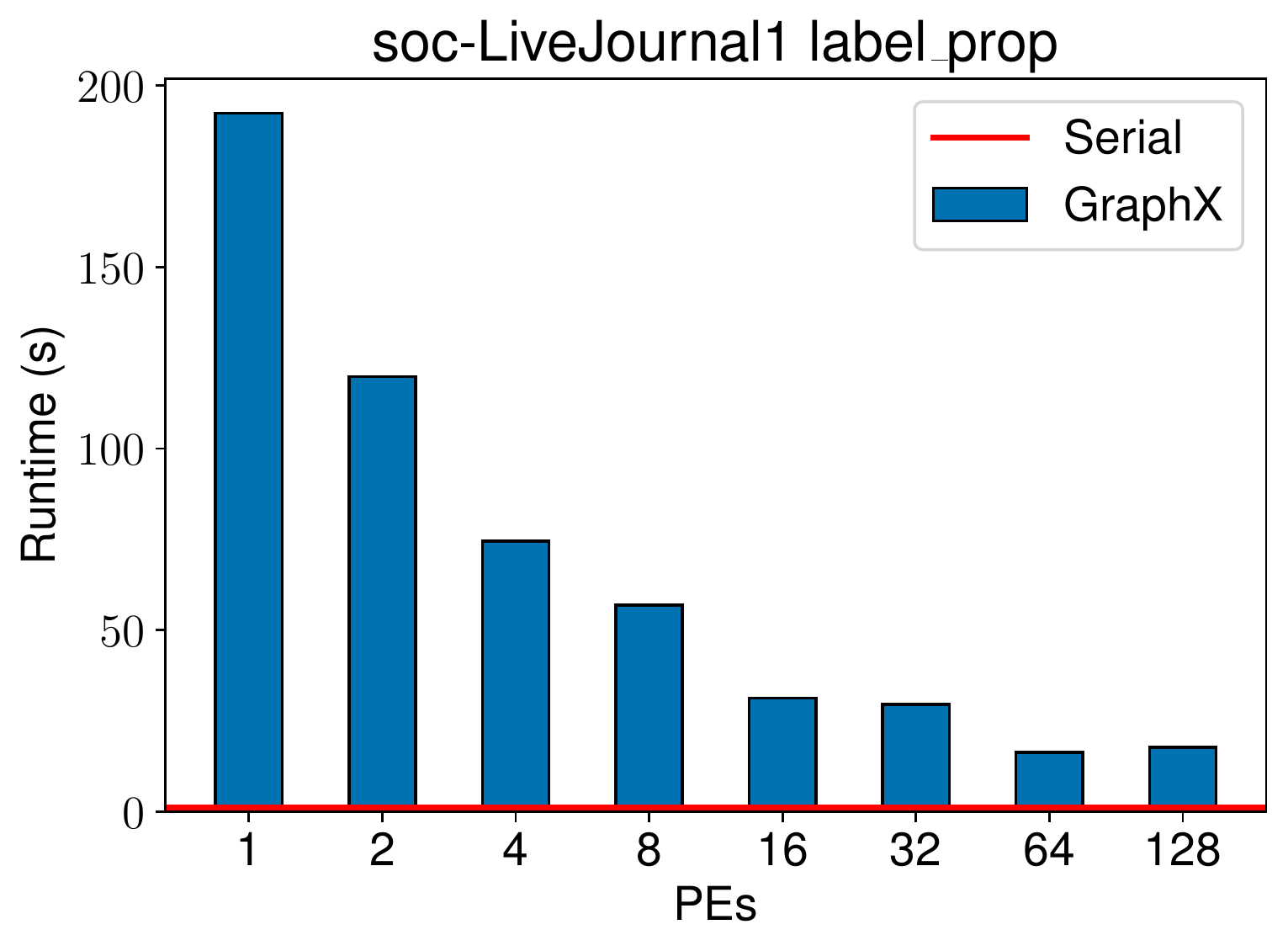}
  \caption{GraphX Label Prop. on \lj{}}
  \label{fig:graphx_cc}
\end{figure}

Figure~\ref{fig:graphx_cc} tells a similar story for label propagation: GraphX
improves with scale, but even its fastest time of 16.31~\si{s} at 64 PEs is over
an order of magnitude slower than the 1.05~\si{s} of the serial version.

These GraphX implementations use GraphX's Pregel API, which claims to avoid
excessive storage of intermediate results for iterative computations like these,
but there appears to be a bug in practice, as even small graphs can cause out of
memory conditions (e.g. a small 11.1 MB test graph running out of memory during
GraphX label propagation given a 16 GB allocation).

Running PageRank with GraphX on \twitter{} also resulted in an OOM crash; the
graph is 5.8 GB and GraphX was given 128 PEs and 180 GB of memory. Running label
propagation on \twitter{} resulted in a timeout after 30 minutes of processing
on 128 PEs. We did not attempt runs with \uk{} due to these failures and the
larger size of that graph.

GraphX was generally outperformed by the serial implementations in the original
COST paper, but not to the extent seen in our experiments. We are using a newer
version of GraphX, which may account for some of the difference, but even so,
the performance gulf is stark. We implemented several different variants of the
GraphX client application in an attempt to improve the performance, but the
results shown were the best we were able to obtain.

These GraphX results correspond to a COST of $\infty$, as performance never
matches the single threaded version, regardless of how much hardware we provide.
We did not perform runs beyond 128 PEs, but scaling improvements appear to end
at or before that point in our results.

\subsection{Charm++}

Our basic Charm++ implementation assigns contiguous chunks of vertices to
chares. Local vertices are stored in index order, and the destinations of
outgoing edges are stored for each vertex. During each iteration, a chare loops
over its local vertices and their outgoing edges, performing any necessary
preparation and aggregating outgoing message data in per-chare buffers. Messages
are sent to their destination chares after the conclusion of this loop. Upon
reception of one of these messages, the chare performs the required computation
to apply the payload of the message to its local vertices.

To analyze their performance impact, we implemented several different variants
with optimizations on top of this basic implementation. Some of these
optimizations violate the semantics of the actor model, namely those of private
state and only exchanging data via messages.

\begin{description}
  \item[Atomic] Passes data using a global vertex buffer updated concurrently via atomic operations instead of using messages.
  \item[Pairs] Passes data using a collection of global buffers, one for each ordered pair of chares instead of using messages. No locks or atomics are needed, synchronized via a message telling consumer that the producer is finished.
  \item[Reduction] Passes data and computes results using a parallel reduction tree instead of using point to point messages. Each chare contributes a buffer containing data for all vertices with the updates coming from its local vertices applied, the buffers are then reduced in parallel, and finally each chare does local updates using the corresponding portion of the reduced buffer.
  \item[Sort Destination] Reorders how edges are stored on a chare; instead of ordering by and storing (local source, \{destinations\}), this orders and stores by (destination, \{list of local sources\}). This orders local iterations by destination, meaning that messages can be sent earlier than in the basic version: after the edges incident to a single destination chunk are done rather than waiting until all edges are done.
\end{description}

\subsubsection{PageRank}

\begin{figure}
  \includegraphics[width=\linewidth]{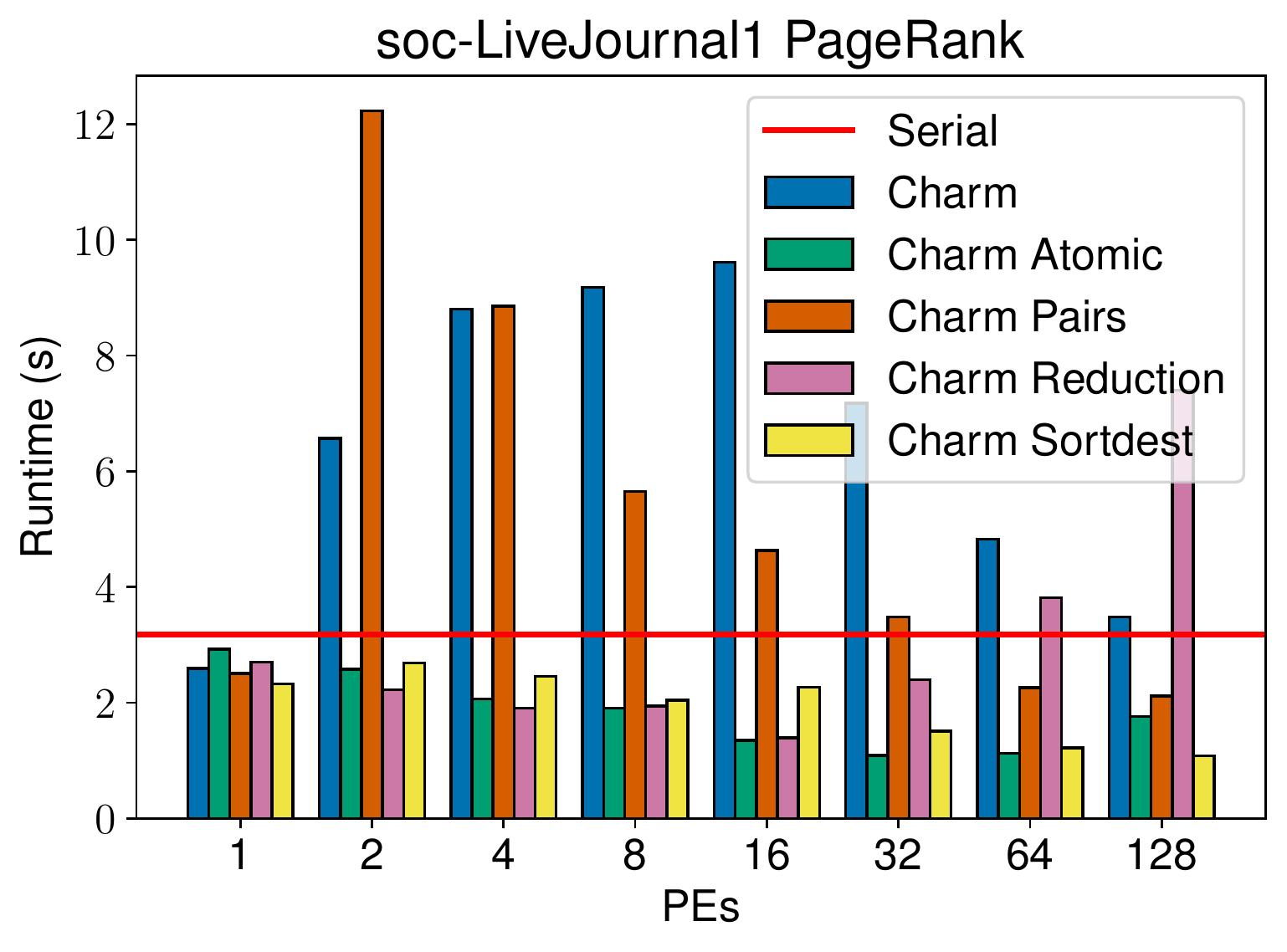}
  \caption{Charm++ PageRank on \lj{}}
  \label{fig:lj_pr}
\end{figure}

\begin{figure}
  \includegraphics[width=\linewidth]{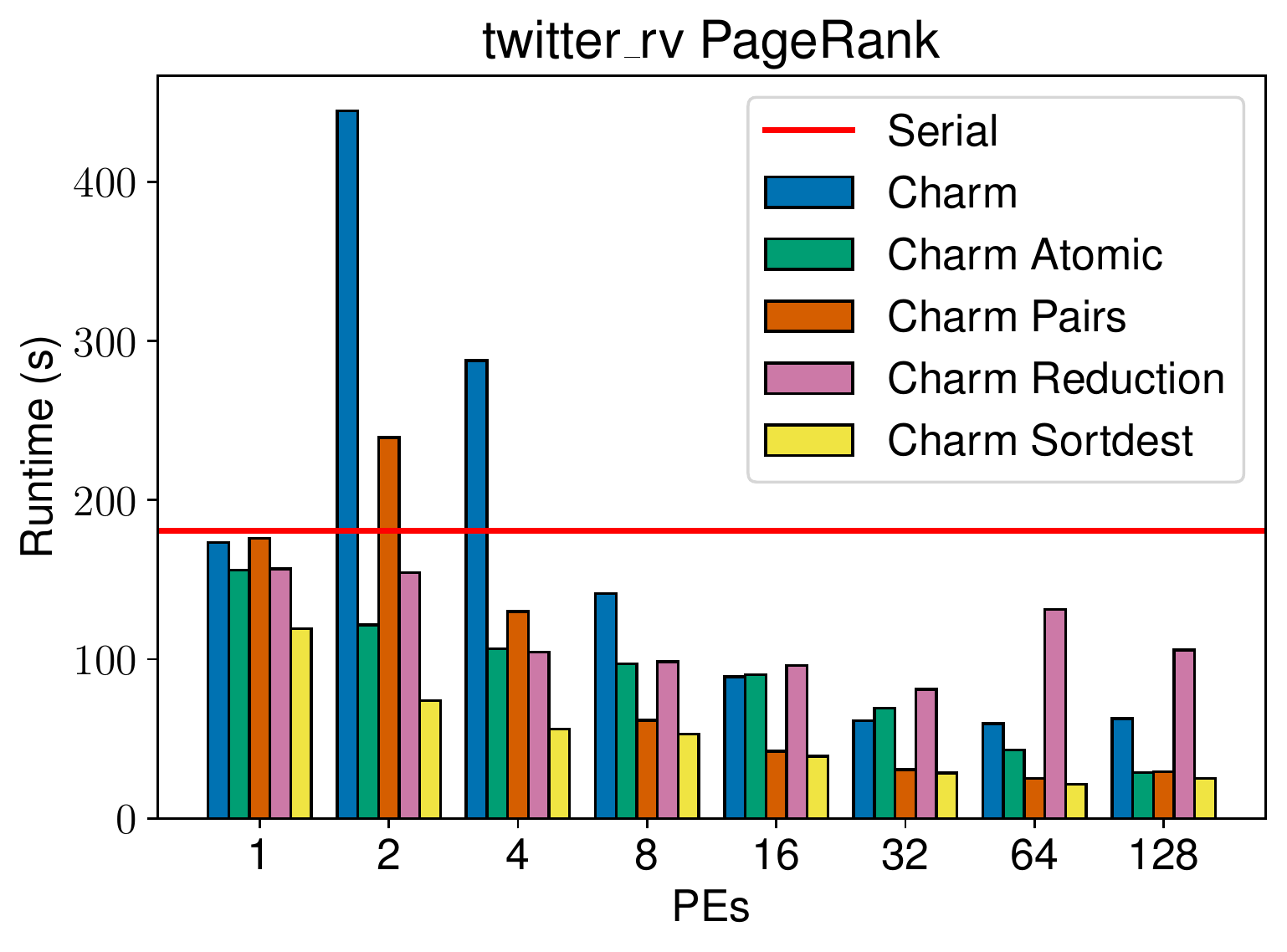}
  \caption{Charm++ PageRank on \twitter{}}
  \label{fig:twitter_pr}
\end{figure}

\begin{figure}
  \includegraphics[width=\linewidth]{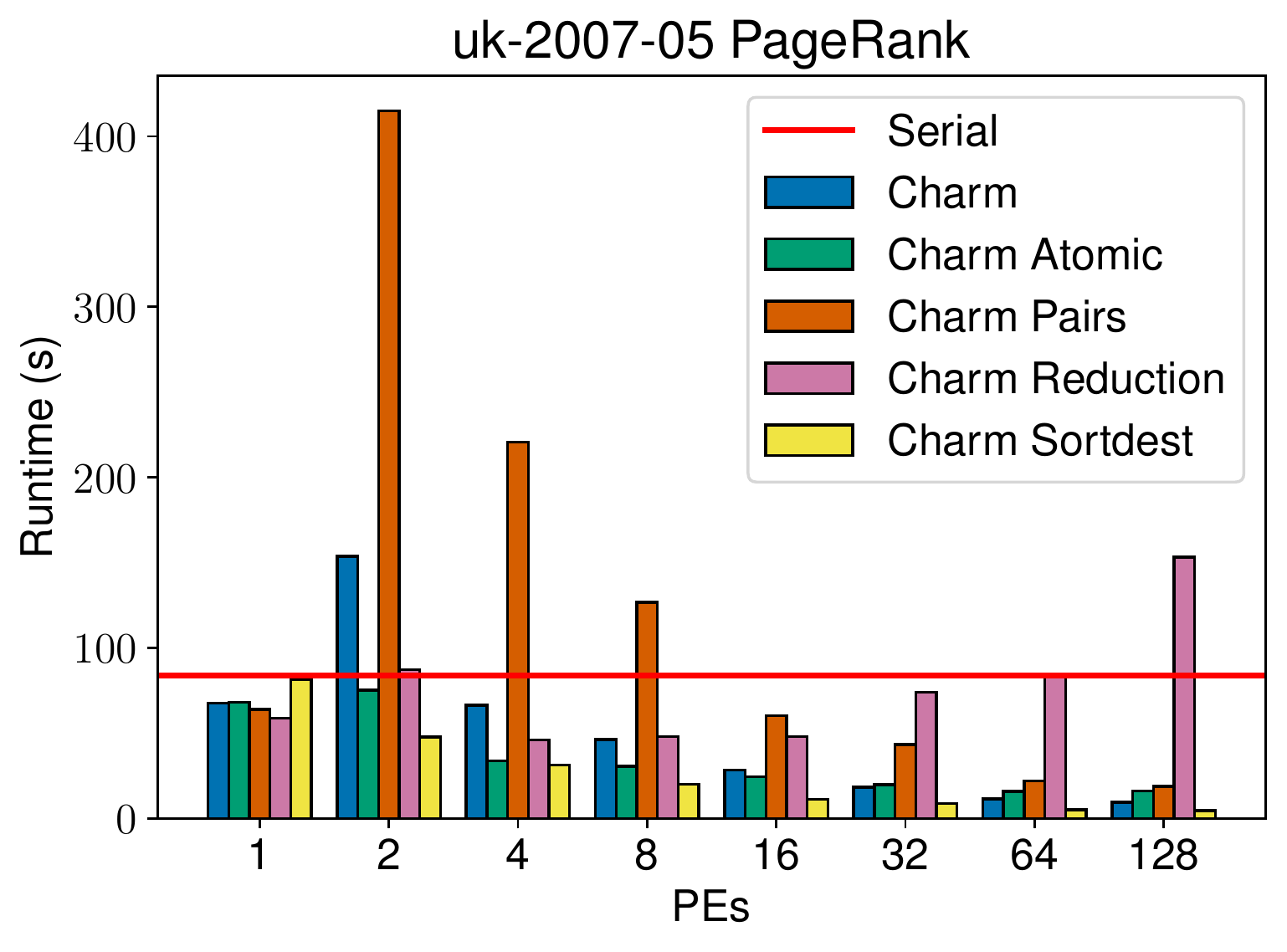}
  \caption{Charm++ PageRank on \uk{}}
  \label{fig:uk_pr}
\end{figure}

\begin{table}[]
  \centering
  \small
  \begin{tabularx}{\linewidth}{lr|rrrl}
    \toprule
    &\multicolumn{1}{l}{}         & \multicolumn{3}{c}{\textbf{Graph}}                                                              &  \\ \cline{3-5}
    &\multicolumn{1}{l}{}         & soc-LJ1 & \twitter{} & \uk{} &  \\
    \midrule
    \multicolumn{2}{c|}{\textbf{Serial}} & 3.18  & 180.69     & 83.62  & \\
    \midrule
    \multirow{8}{*}{\textbf{PEs}} &  1                        & 2.33                      & 119.28                         & 58.65                     &  \\
    &2                            & 2.22                      & 73.93                          & 47.83                     &  \\
    &4                            & 1.90                      & 56.01                          & 31.36                     &  \\
    &8                            & 1.90                      & 52.84                          & 19.94                     &  \\
    &16                           & 1.35                      & 39.02                          & 11.19                     &  \\
    &32                           & 1.09                      & 28.55                          & 8.61                      &  \\
    &64                           & 1.13                      & 21.48                          & 5.11                      &  \\
    &128                          & 1.08                      & 25.00                          & 4.56                      & \\
    \bottomrule
  \end{tabularx}
  \caption{Best Charm++ PageRank Runtime (s) Across All Variants}
  \label{tab:pr_perf}
\end{table}

Figures~\ref{fig:lj_pr},~\ref{fig:twitter_pr}, and~\ref{fig:uk_pr} show PageRank
performance for the Charm++ variants and the serial implementation on \lj{},
\twitter{}, and \uk{}, respectively. Table~\ref{tab:pr_perf} shows the runtime
of the best performing variant at every scale.

Significantly, the COST with all input graphs is 1, as the performance of the
Charm++ variants matches or exceeds the performance of the serial version on a
single processor. However, the performance and scalability of the different
variants varies greatly.

The basic variant performs well on 1 PE, but then spikes in runtime when moving
to 2 PEs, before regaining performance with scale. This is due to the allocation
and serialization overhead of messaging, which diminishes in relative importance
as buffers become smaller as the computation is scaled.

The atomic variant generally performs well at all scales and for all graphs,
since it avoids the overheads of messaging and uses a highly performant
technique to operate on shared data. While the use of shared state breaks a
tenant of the actor model and does not work for distributed execution, from a
programmabilty perspective, using atomics is simpler and less error-prone than
using locks while also being suitable for fine-grained parallelism.

The pairs variant works fairly well at the medium to large scale on \twitter{}
but poorly on the other graphs and at smaller scale. As with the basic variant,
this is likely due to allocation overheads from needing to dynamically manage
multiple large buffers, possibly exacerbated by NUMA effects. Note that the
buffers here are proportional in size to the number of edges, rather than the
much smaller number of vertices, as in the atomic variant. This variant shows
one of the risks of using shared state: getting a pointer to it may be cheap,
but managing it may be costly.

The reduction variant is very reasonable at small to medium scale, but is the
worst performing variant for every graph at large scale. Its poor performance is
due to two factors: load imbalance and memory utilization.

\begin{enumerate}
  \item In the other variants (with the exception of atomic, which has the same
        global synchronization), a chare can progressively proceed with applying
        updates as neighboring chares send data to it, whereas the reduction
        variant requires all chares to have finished their local loop, as the
        reduction can only complete after all objects have contributed their
        data.
  \item Each chare must allocate a buffer of size equal to the number of total
        vertices in the graph to contribute to the reduction since sparse
        contributions are not supported. The other variants communicate using
        space proportional to the number of edges (with the exception of atomic,
        which uses a single global vertex buffer). While the number of edges is
        larger than the number of vertices for all of the input graphs, it is
        only 14-35x the number of vertices, meaning the reduction variant will
        be using much more memory at 128 PEs, leading to allocation overhead and
        more cache evictions.
\end{enumerate}

The sort destination variant is the best performing overall, due to
several different reasons:

\begin{enumerate}
  \item Better locality and more efficient memory utilization, as it only needs
        to maintain a single buffer for message data since it computes all of
        the outgoing data for a single destination chunk before moving onto the
        next chunk. Similarly, on the receive side, the payload of the message
        is arranged in the same order as the local vertex data are stored.
  \item Reduced load imbalance by sending messages earlier than other variants,
        allowing otherwise idle chares to move onto the next phase of the
        iteration instead of waiting with no work to do.
  \item Sending less data than other variants by locally reducing the data bound
        for an external vertex before sending, which is also enabled by its
        access pattern. As an example, suppose two vertices A and B on a chare
        both have outgoing edges to an external vertex C. In the basic variant,
        the update from A is added to a message buffer, and later the update
        from B is added to the same message buffer. We could search this buffer
        to combine the two updates, but doing so is relatively costly, requiring
        either a linear search or maintaining some sorted or hash-based data
        structure and the associated memory and computational overheads, so we
        send the update from A and B separately. On the other hand, due to the
        arrangement of edges in the sort destination variant, we process all
        vertices with an outgoing edge to C consecutively, so combining the
        update from A with the update from B before sending is trivial.
\end{enumerate}

\subsubsection{Label Propagation}

\begin{figure}
  \includegraphics[width=\linewidth]{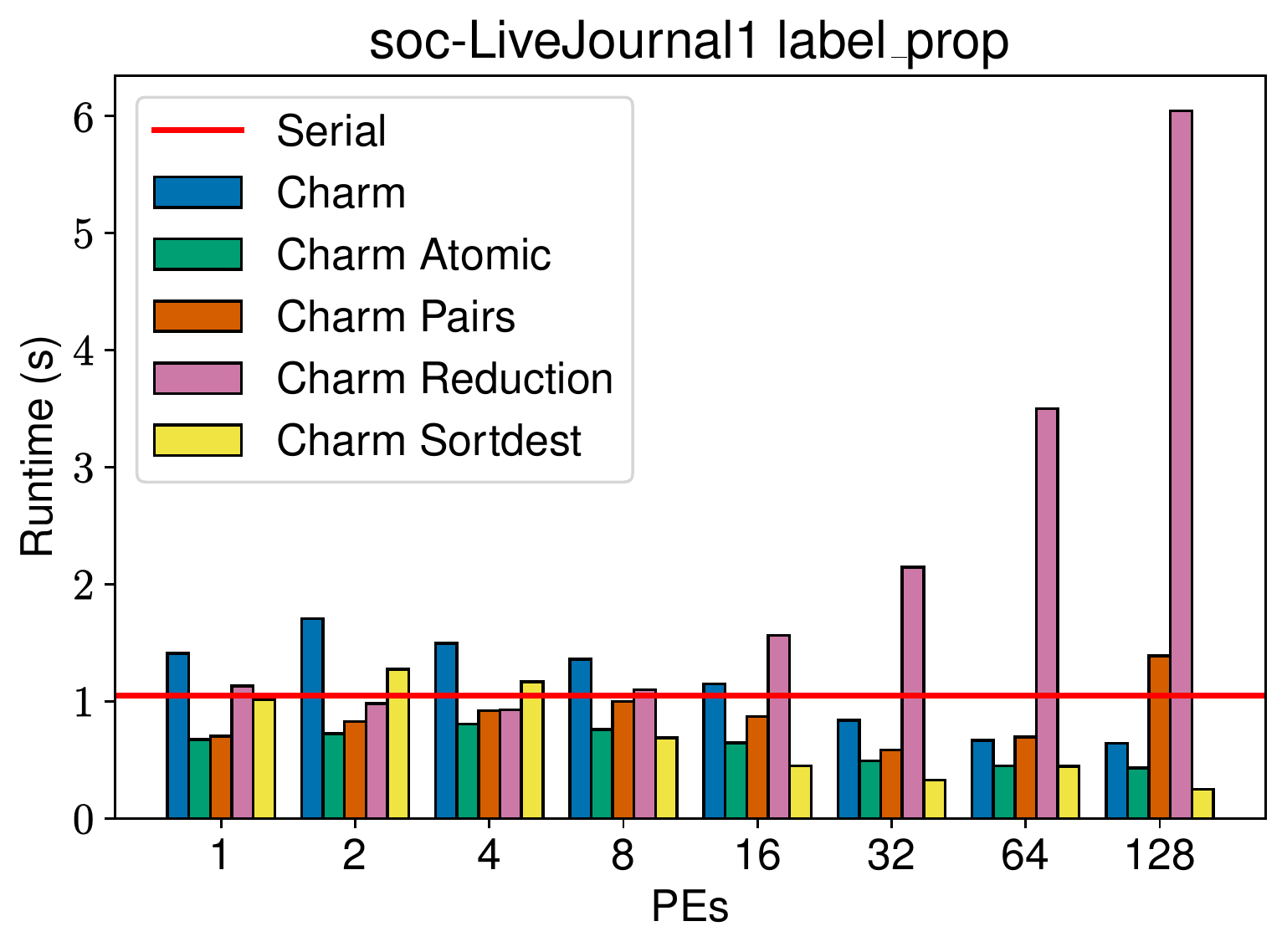}
  \caption{Charm++ Label Prop. on \lj{}}
  \label{fig:lj_cc}
\end{figure}

\begin{figure}
  \includegraphics[width=\linewidth]{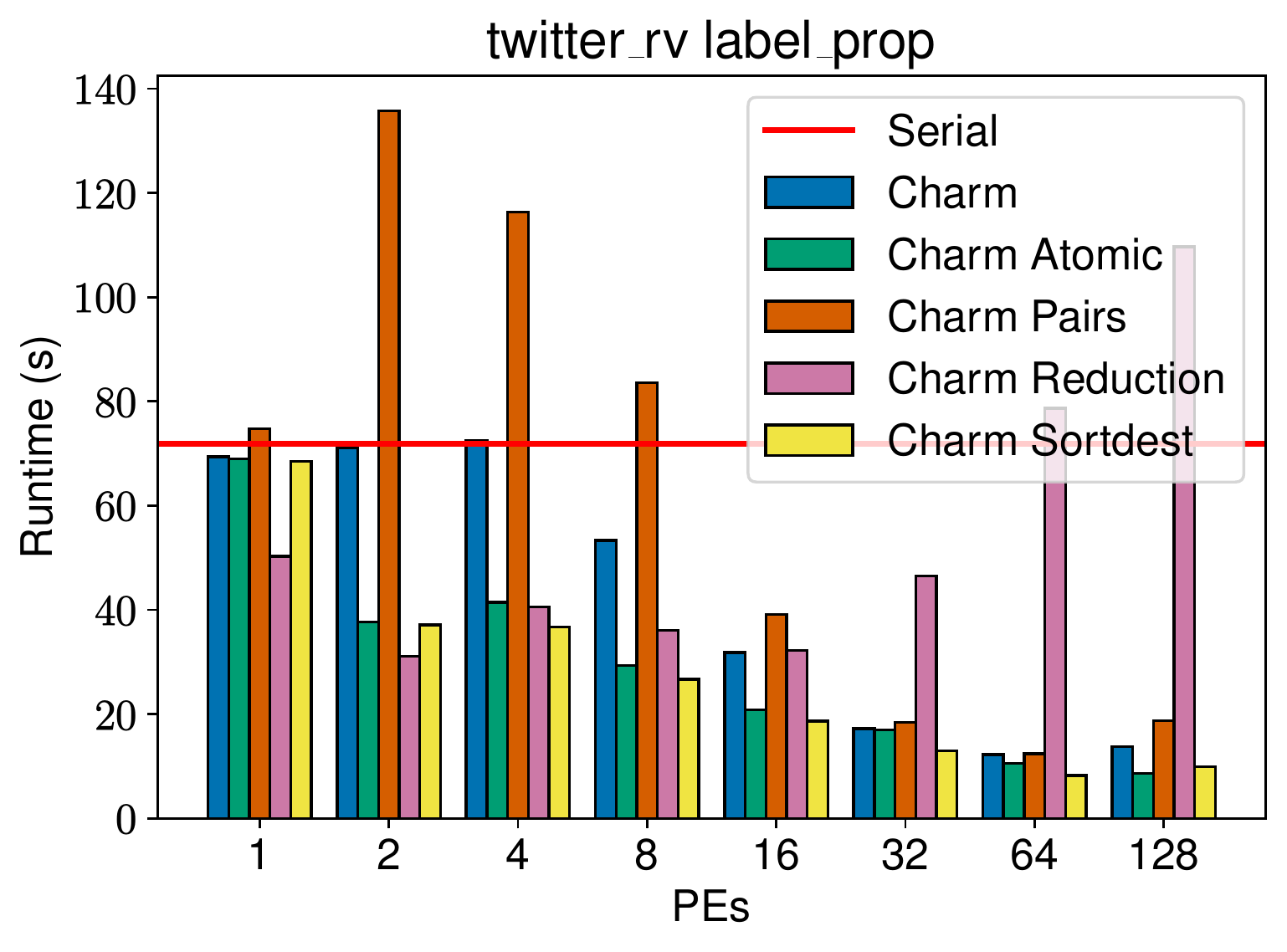}
  \caption{Charm++ Label Prop. on \twitter{}}
  \label{fig:twitter_cc}
\end{figure}

\begin{figure}
  \includegraphics[width=\linewidth]{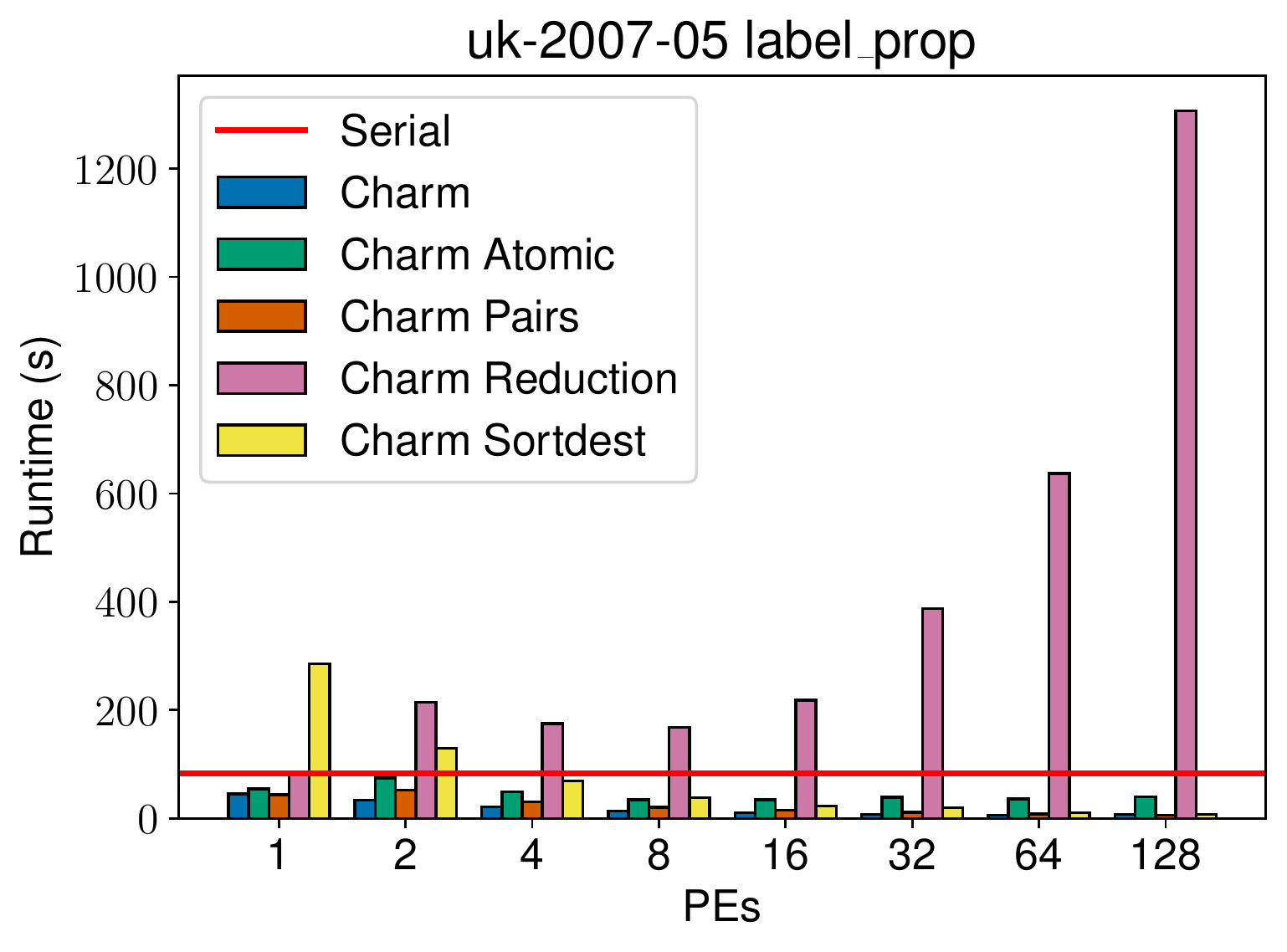}
  \caption{Charm++ Label Prop. on \uk{}}
  \label{fig:uk_cc}
\end{figure}

\begin{table}[]
  \centering
  \small
  \begin{tabularx}{\linewidth}{lr|rrr}
    \toprule
    &\multicolumn{1}{l}{}         & \multicolumn{3}{c}{\textbf{Graph}}                                                                \\ \cline{3-5}
    &\multicolumn{1}{l}{}         & soc-LJ1 & \twitter{} & \uk{} \\
    \midrule

    \multicolumn{2}{c|}{\textbf{Serial}} & 1.05  & 71.85                          & 83.59 \\
    \midrule
    \multirow{8}{*}{\textbf{PEs}} & 1                            & 0.67                      & 50.32                          & 43.83                     \\
    & 2                            & 0.72                      & 31.07                          & 33.45                     \\
    & 4                            & 0.80                      & 36.68                          & 21.56                     \\
    & 8                            & 0.69                      & 26.69                          & 13.81                     \\
    & 16                           & 0.45                      & 18.68                          & 10.24                     \\
    & 32                           & 0.33                      & 12.97                          & 7.15                      \\
    & 64                           & 0.45                      & 8.24                           & 5.69                      \\
    & 128                          & 0.25                      & 8.66                           & 6.11 \\
    \bottomrule
  \end{tabularx}
  \caption{Best Charm++ Label Propagation Runtime (s) Across All Variants}
  \label{tab:cc_perf}
\end{table}

Figures~\ref{fig:lj_cc},~\ref{fig:twitter_cc}, and~\ref{fig:uk_cc} show label
propagation performance for the Charm++ variants and the serial implementation
on \lj{}, \twitter{}, and \uk{}, respectively. Table~\ref{tab:cc_perf} shows the
runtime of the best performing variant at every scale.

Scaling performance for the various variants with label propagation are similar
to those of PageRank. At least one variant is faster than the serial baseline
version for every graph and at every scale, so the COST is again 1.

One notable difference between the PageRank and label propagation
implementations is that PageRank sends the same volume of data at every
iteration, whereas label propagation sends a variable amount of data, only
sending updates on edges coming from a vertex that has updated its label since
the previous iteration (does not apply to the reduction variant, which continues
sending a buffer of global vertices). The serial implementation does not have
this optimization

The reduction variant is even slower at the large scale here than it was for
PageRank. Structurally, the computation is not very different, so the
performance delta is likely due to increased cache or memory pressure because of
the larger number of edges used for label propagation.

\section{Conclusions}
\label{sec:conclusions}

Using COST is a simple, practical way to comparatively assess the performance of
concurrent systems. Too often, benchmarks are given in isolation without
providing a reasonable baseline, highlighting specious scalability over
practical performance.

Developing performant, scalable software is difficult and fraught with the
complexities of managing threads, messaging, synchronization, scheduling, and
more. The actor model provides an elegant way to design concurrent applications,
ameliorating many of the traditional difficulties of parallel programming
without adding constraining restrictions or sacrificing performance.

It is a testament to the actor model that with only minor changes to convert a
simple serial code into a message driven, actor-based, concurrent
implementation, we were able to achieve scalability in parallel execution while
maintaining absolute performance, matching the serial version when executing on
a single processor. Furthermore, our implementations greatly outperformed a
purportedly scalable ``big data'' system, providing scalable performance up to
128 PEs.

Optimizations that violated the semantics of the pure actor model by using
shared state were helpful in some cases. However, empirically, the most
beneficial optimization, sort destination, merely involved altering the order
and organization of private data within a chare to improve cache performance,
reduce load imbalance, and shrink messages sizes by doing local reductions.

\section{Acknowledgements}

This research used the Delta advanced computing and data resource which is
supported by the National Science Foundation (award OAC 2005572) and the State
of Illinois. Delta is a joint effort of the University of Illinois
Urbana-Champaign and its National Center for Supercomputing Applications.


\clearpage
\printbibliography

@inproceedings{McSherryCOST,
author = {McSherry, Frank and Isard, Michael and Murray, Derek G.},
    title = {Scalability! But at What Cost?},
    year = {2015},
    publisher = {USENIX Association},
    address = {USA},
    booktitle = {Proceedings of the 15th USENIX Conference on Hot Topics in Operating Systems},
    pages = {14},
    numpages = {1},
    location = {Switzerland},
    series = {HOTOS'15}
    }

@misc{snapnets,
  author       = {Jure Leskovec and Andrej Krevl},
  title        = {{SNAP Datasets}: {Stanford} Large Network Dataset Collection},
  howpublished = {\url{http://snap.stanford.edu/data}},
  month        = jun,
  year         = 2014
}

@inproceedings{Kwak10www,
author = {Kwak, Haewoon and Lee, Changhyun and Park, Hosung and Moon, Sue},
title = "{W}hat is {T}witter, a social network or a news media?",
booktitle = {WWW '10: Proceedings of the 19th international conference on World wide web},
year = {2010},
isbn = {978-1-60558-799-8},
pages = {591--600},
location = {Raleigh, North Carolina, USA},
doi = {http://doi.acm.org/10.1145/1772690.1772751},
publisher = {ACM},
address = {New York, NY, USA},
}

@inproceedings{BoVWFI,
  author = "Paolo Boldi and Sebastiano Vigna",
  title = "The {W}eb{G}raph Framework {I}: {C}ompression Techniques",
  year = 2004,
  booktitle = "Proc. of the Thirteenth International World Wide Web Conference (WWW 2004)",
  address = "Manhattan, USA",
  pages = "595--601",
  publisher = "ACM Press"
}

@inproceedings{BRSLLP,
  author = "Paolo Boldi and Marco Rosa and Massimo Santini and Sebastiano Vigna",
  title = "Layered Label Propagation: A MultiResolution Coordinate-Free Ordering for Compressing Social Networks",
  booktitle = "Proceedings of the 20th international conference on World Wide Web",
  editor = "Sadagopan Srinivasan and Krithi Ramamritham and Arun Kumar and M. P. Ravindra and Elisa Bertino and Ravi Kumar",
  publisher = "ACM Press",
  year = 2011,
  pages = "587--596"
}

@techreport{pagerank,
          number = {1999-66},
           month = {November},
          author = {Lawrence Page and Sergey Brin and Rajeev Motwani and Terry Winograd},
            note = {Previous number = SIDL-WP-1999-0120},
           title = {The PageRank Citation Ranking: Bringing Order to the Web.},
            type = {Technical Report},
       publisher = {Stanford InfoLab},
            year = {1999},
     institution = {Stanford InfoLab},
             url = {http://ilpubs.stanford.edu:8090/422/}
}

@inproceedings{graphx,
author = {Gonzalez, Joseph E. and Xin, Reynold S. and Dave, Ankur and Crankshaw, Daniel and Franklin, Michael J. and Stoica, Ion},
title = {GraphX: Graph Processing in a Distributed Dataflow Framework},
year = {2014},
isbn = {9781931971164},
publisher = {USENIX Association},
address = {USA},
booktitle = {Proceedings of the 11th USENIX Conference on Operating Systems Design and Implementation},
pages = {599–613},
numpages = {15},
location = {Broomfield, CO},
series = {OSDI'14}
}

@inproceedings{actor,
author = {Hewitt, Carl and Bishop, Peter and Steiger, Richard},
title = {A Universal Modular ACTOR Formalism for Artificial Intelligence},
year = {1973},
publisher = {Morgan Kaufmann Publishers Inc.},
address = {San Francisco, CA, USA},
booktitle = {Proceedings of the 3rd International Joint Conference on Artificial Intelligence},
pages = {235–245},
numpages = {11},
location = {Stanford, USA},
series = {IJCAI'73}
}

@misc{charm-zenodo,
    author = {Kale, Laxmikant and Acun, Bilge and Bak, Seonmyeong and Becker, Aaron and
      Bhandarkar, Milind and Bhat, Nitin and Bhatele, Abhinav and Bohm, Eric and
      Bordage, Cyril and Brunner, Robert and Buch, Ronak and Chakravorty, Sayantan
      and Chandrasekar, Kavitha and Choi, Jaemin and Denardo, Michael and DeSouza,
      Jayant and Diener, Matthias and Dokania, Harshit and Dooley, Isaac and Fenton,
      Wayne and Galvez, Juan and Gioachin, Fillipo and Gupta, Abhishek and Gupta,
      Gagan and Gupta, Manish and Gursoy, Attila and Harsh, Vipul and Hu, Fang and
      Huang, Chao and Jagathesan, Narain and Jain, Nikhil and Jetley, Pritish and
      Jindal, Prateek and Kanakagiri, Raghavendra and Koenig, Greg and Krishnan,
      Sanjeev and Kumar, Sameer and Kunzman, David and Lang, Michael and Langer,
      Akhil and Lawlor, Orion and Wai Lee, Chee and Lifflander, Jonathan and Mahesh,
      Karthik and Mendes, Celso and Menon, Harshitha and Mei, Chao and Meneses,
      Esteban and Mikida, Eric and Miller, Phil and Mokos, Ryan and Narayanan,
      Venkatasubrahmanian and Ni, Xiang and Nomura, Kevin and Paranjpye, Sameer and
      Ramachandran, Parthasarathy and Ramkumar, Balkrishna and Ramos, Evan and
      Robson, Michael and Saboo, Neelam and Saletore, Vikram and Sarood, Osman and
      Senthil, Karthik and Shah, Nimish and Shu, Wennie and B. Sinha, Amitabh and
      Sun, Yanhua and Sura, Zehra and Totoni, Ehsan and Varadarajan, Krishnan and
      Venkataraman, Ramprasad and Wang, Jackie and Wesolowski, Lukasz and White, Sam
      and Wilmarth, Terry and Wright, Jeff and Yelon, Joshua and Zheng, Gengbin},
    doi = {10.5281/zenodo.3370873},
    month = {Aug},
    title = {{The Charm++ Parallel Programming System}},
    url = {https://charm.cs.illinois.edu},
    year = {2019},
  }

@incollection{Kale2013,
        address = {Boca Raton, FL, USA},
        author = {Kale, Laxmikant V. and Zheng, Gengbin},
        booktitle = {{Parallel Science and Engineering Applications: The Charm++ Approach}},
        chapter = 1,
        doi = {10.1201/b16251},
        edition = {1st},
        editor = {Kale, Laxmikant V. and Bhatele, Abhinav},
        isbn = {1466504129, 9781466504127},
        pages = {1-16},
        publisher = {CRC Press, Inc.},
        title = {{Chapter 1: The Charm++ Programming Model}},
        year = {2013},
      }

@article{pregel,
author = {Yan, Da and Cheng, James and Xing, Kai and Lu, Yi and Ng, Wilfred and Bu, Yingyi},
title = {Pregel Algorithms for Graph Connectivity Problems with Performance Guarantees},
year = {2014},
issue_date = {October 2014},
publisher = {VLDB Endowment},
volume = {7},
number = {14},
issn = {2150-8097},
url = {https://doi.org/10.14778/2733085.2733089},
doi = {10.14778/2733085.2733089},
journal = {Proc. VLDB Endow.},
month = {oct},
pages = {1821–1832},
numpages = {12}
}


\end{document}